\documentclass[12pt,a4paper,final]{iopart}
\usepackage{iopams}
\usepackage{graphicx}
\usepackage{epstopdf}
\usepackage[breaklinks=true,colorlinks=true,linkcolor=blue,urlcolor=blue,citecolor=blue]{hyperref}

\begin{document}

\title[]{Electric field-induced chiral $d+id$ superconducting state in AA-stacked bilayer graphene: A quantum Monte Carlo study}

\author{Shi-Chao Fang$^{1}$, Yan Zhang$^{2}$, Xiaojun Zheng$^3$, Guangkun Liu$^{4}$, Zhongbing Huang$^{1,4,*}$}
\address{$^1$Faculty of Physics and Electronic Technology, Hubei University, Wuhan 430062, China}
\address{$^2$College of Physics and Electronic Engineering, Northwest Normal University, Lanzhou 730070, China }
\address{$^3$College of Science, Guilin University of Technology, Guilin 541004, China}
\address{$^4$Beijing Computational Science Research Center, Beijing 100193, China}

\ead{huangzb@hubu.edu.cn}

\begin{abstract}
Using constrained-path quantum Monte Carlo method, we systematically study the
Hubbard model on AA-stacked honeycomb lattices with electric field. Our simulation demonstrates a dominant chiral $d+id$ wave pairing induced by the electric field at half filling. In particular, as the on-site Coulomb interaction increases, the effective pairing correlation of chiral $d+id$ superconducting state exhibits increasing behavior. We attribute the electric field induced $d+id$ superconductivity to an increased density of states near the Fermi energy and an suppressed antiferromagnetic spin correlation after turning on the electric field. Our results strongly suggest the AA-stacked graphene system with electric field is a good candidate for chiral $d+id$ superconductors.
\end{abstract}


\section{Introduction}
  Unconventional superconductivity continues to attract the attention of condensed
matter community~\cite{Dagotto2013,Johnston2010,Mackenzie2003,Scalapino2012,Dagotto1994,Qi2011,Mazin2008}.
Recently, for the search on unconventional superconducting states, graphene with special electronic structure has become a more concerned object \cite{Castro2009}. Theoretically, single-layer graphene exhibits a rich superconducting phase diagram. For the doping near the half-filling and the van Hove singularities, it is generally believed that there is a $d+id$ (i.e., $d_{x^{2}-y^{2}}+id_{xy}$) superconducting state \cite{Black2012,Black2014,Nandkishore2012,Nandkishore2014,Pathak2010,
Ma2011,Ying2018,Black-Schaffer2007}. However, recent theoretical results show that there may be a coexistence of chiral $d+id$ wave pairing and triplet $f$ wave pairing in the low doping region \cite{Kiesel2012,Honerkamp2008,Raghu2010}. For the deeply doped region, the triplet $p+ip$ (i.e., $p_{x}+ip_{y}$) pairing is widely predicted by theoretical calculations in single layer graphene systems \cite{Ma2014,Bernardo2017,Ma2017,Faye2015,Xiao2015,Xu2016}. The chiral $d+id$ superconducting state is the most interesting for the graphene systems. Firstly, the chiral $d+id$ superconducting state breaks the time-reversal symmetry, which can  exhibit non-trivial topological properties \cite{Vojta2000}, such as the quantized spin Hall conductivity and the thermal Hall conductivity \cite{Moore1991}, the Mayorana state in the superconducting vortex \cite{Kopnin1991}, and the gapless state on the boundary \cite{Volovik1989}.
Secondly, the symmetry of graphene ensures that $d_{x^{2}-y^{2}}$ and $d_{xy}$ superconducting pairing channels are degenerate \cite{Black-Schaffer2007,Xu2016,Gonzalez2008}, which makes graphene  a potential candidate for the formation of chiral $d+id$ superconducting state.

More recently, the discovery of correlated-insulating and superconducting states in twisted bilayer graphene (TBG) has attracted unprecedented attention~\cite{Cao2018a,Cao2018b}. The experimental results  show that the TBG exhibits insulator behavior at half-filling \cite{Cao2018a}, and doping can induce the transformation from insulator to superconductor at $1.7$ K \cite{Cao2018b}. Nevertheless, the pairing of TBG is still a controversial topic \cite{Isobe2018,Gonzalez2018,Liu2018,Xu2018,Kennes2018,Peltonen2018,Fidrysiak2018}. One inherent advantage of two-dimensional materials is that the chemical potential of electrons can be constantly adjusted through an electric field without introducing additional disorder. Excitedly, recent experiments have reported the superconductivity with critical transition temperature of $12$ K in bilayer graphene with a twist angle of $1.28^\circ$ by electric field control \cite{Shen2019}. Inspired by this, we studied the superconducting pairing by adjusting the electric field  in the AA stacked bilayer graphene.

AA-stacked bilayer graphene~\cite{Liu2009} is the simplest form among bilayer graphene systems, which has attracted extensive attentions
~\cite{Liu2009,Rozhkov2016,Rakhmanov2012,Andres2008,Akzyanov2014,Sboychakov2013a,Sboychakov2013b}.
Some works discussed the magnetic properties of the AA-stacked bilayer. For
instances, Akzyanov et al studied the electronic properties
of AA-stacked bilayer graphene and found that antiferromagnetic order is suppressed by the transverse electric field \cite{Akzyanov2014}.
Sboychakov et al. \cite{Sboychakov2013a,Sboychakov2013b} found that the AA-stacked system is
an antiferromagnetic insulator at half filling and the slim doping could induce a possible
metal-insulator transition. However, few works have been published in the field of superconductivity in the AA-stacked bilayer graphene.

In this paper, we studied the electron pairings on the AA-stacked bilayer graphene
system. By using constrained-path quantum Monte Carlo method (CPQMC)
\cite{Zhang1997a,Zhang1997b,Huang2001a,Huang2001b,Liu2014,Liu2016}, we systematically
studied the impact of electric field on the AA-stacked bilayer graphene and
our simulation results demonstrate a dominant chiral $d+id$-wave pairing triggered by the electric field. We find that such a dominance of chiral $d+id$-wave pairing comes from the significantly
increased  density of states near Ferimi level and also enhanced
 antiferromagnetic fluctuations by the electric field.

The organization of this publication is as follows: the model and the Monte Carlo method we used are described in Section 2.  Section 3 contains our main numerical
results, and finally in Section 4 we provide our conclusions.

\section{Model and method}
\begin{figure}
\centering
\includegraphics[scale=0.7]{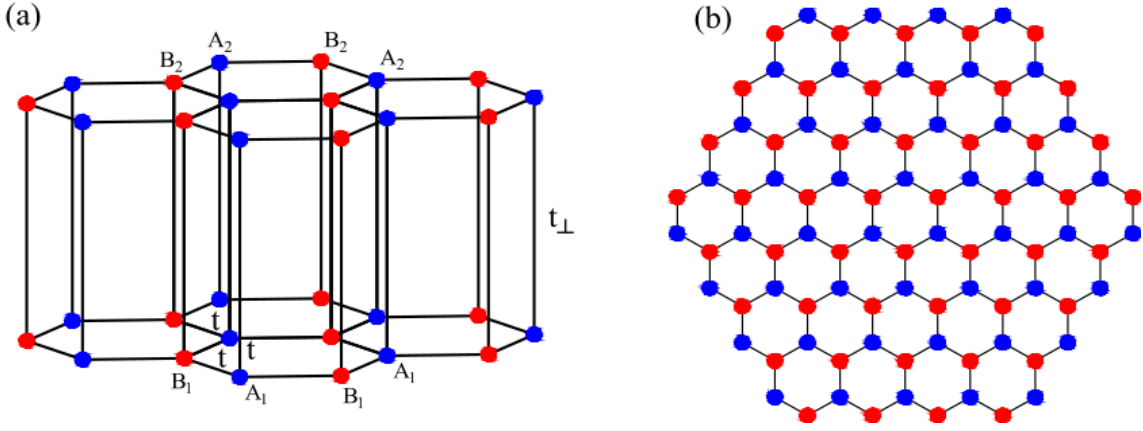}
\caption{(color online)
(a) Sketch of an AA-stacked honeycomb lattice. Blue (red) dots represent sublattice
A (B). A$_1$,B$_1$ (A$_2$,B$_2$) represent sublattices in layer 1 (layer 2).
t and $t_{\perp}$ are the intralayer and interlayer hoppings, respectively.
(b) Geometry of each graphene
layer. Atom number on each layer is $2\times3L^2$. The presenting lattice is corresponding
to $L=4$.
}\label{lattice}
\end{figure}
We study the Hubbard model with an electric field on the AA-stacked bilayer
honeycomb lattices. The sketch of AA-stacked honeycomb lattice
is shown in Figure ~\ref{lattice}(a). The Hubbard model can be described as follows,

\begin{eqnarray}
H&=&-t\sum_{\langle i,j\rangle,m,\sigma}
(c_{i,m,\sigma}^{\dagger}c_{j,m,\sigma}+ \mathrm{h.c.})
-t_{\perp}\sum_{i,\sigma}
(c_{i,1,\sigma}^{\dagger}c_{i,2,\sigma} + \mathrm{h.c.})
\nonumber\\
& &+U\sum_{i,m}n_{i,m,\uparrow}n_{i,m,\downarrow}
 +\varepsilon\sum_{i\sigma}c_{i,2,\sigma}^{\dagger}c_{i,2,\sigma},
\end{eqnarray}
where $c_{i,m,\sigma}^{\dagger}(c_{i,m,\sigma})$ creates (annihilates) an
electron at site $i$ of layer $m$ ($m=1,2$) with spin
$\sigma$ $(\sigma=\uparrow, \downarrow)$.  Hopping $t$ ($t_{\perp}$) connects
intralayer (interlayer) nearest-neighbor sites. According to ab initio calculations and experiments \cite{Castro2009,Rozhkov2016,Charlier1992,Dresselhaus2002}, the realistic value of intralayer nearest-neighbor hopping integral $t=2.5-3$ eV and the interlayer nearest-neighbor hopping integral $t_{\perp}=0.3-0.4$ eV. For simplicity, we unified $t$ as $t=1.0$ and kept $t_{\perp}=0.14t$ in later simulations. $n_{i,m,\sigma}$ is the electron number operator and $U$ denotes the on-site Hubbard repulsive interaction. In order to apply the perpendicular electric
filed to the bilayer system, we put potential difference $\varepsilon$ between the two layers. The current experimental work on bilayer graphene can tune the potential difference up to about $1.0$ eV.  \cite{Shen2019,Zhang2009,Castro 2007},  For this reason, we selected  $0.0t-1.0t$ as the parameter range of potential difference $\varepsilon$.

\begin{figure}
\centering
\includegraphics[scale=0.7]{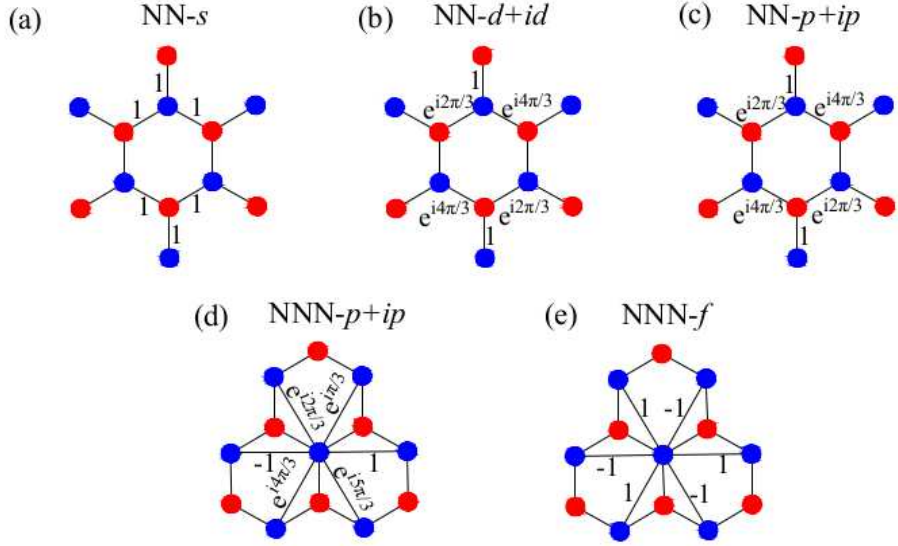}
\caption{(color online)
Sketch of calculated intralayer pairing channels. (a) nearest-neighbor (NN) $s$-wave
(b) NN $d+id$-wave (c) NN $p+ip$-wave  (d) next-nearest-neighbor (NNN) $p+ip$
wave (e) NNN $f$-wave pairings. }\label{pairingstructure}
\end{figure}

Since the atom number in each layer equals to $2\times3L^2$, as shown in Figure ~\ref{lattice}(b), the total lattice number is equal to $4\times3L^2$.
Our main numerical calculations are performed on $L=4, 5, 6$ systems with periodic boundary condition on each layer.
To investigate the pairing properties, we calculate the pairing correlations for various pairing channels in the same layer.
Pairing correlation function can be defined as
\begin{equation}
C_{\alpha}(R=|i-j|)=\langle\Delta_{\alpha}^{\dagger}(i)\Delta_{\alpha}(j)\rangle,
\end{equation}
where $\Delta_{\alpha}^{\dagger}(i)$ ($\Delta_{\alpha}(i)$) is the electron pair creation (annihilation) operator with pairing symmetry $\alpha$. Singlet or triplet pair creation operator can be written as,

\begin{equation}
\Delta_{\alpha}^{\dagger}(i) =\frac{1}{\sqrt{N_{\alpha}}} \sum_l f^{\dagger}_{\alpha}(\delta_{l})(
c_{i,\uparrow}c_{i+\delta_l,\downarrow}\mp
c_{i,\downarrow}c_{i+\delta_l,\uparrow})^{\dagger},
\end{equation}
where $f_{\alpha}(\delta_l)$ is the form factor distinguishing different
pairing channels. -(+) corresponds to single (triplet) pairing. The vectors $\delta_l$ denote the nearest-neighbor (NN) inter-sublattice  or
next-nearest-neighbor (NNN) intra-sublattice connections. $N_{\alpha}$ are the corresponding normalization factors with $N_{\alpha}=3$ ($N_{\alpha}=6$) for NN (NNN) channels. Here we considered the
common NN and NNN pairings, such as
NN  $s$-wave, NN $d +id$-wave, and NN $p + ip$-wave, NNN
$p + ip$-wave, and NNN $f$-wave pairings symmetries. The factors
$f_{\alpha}(\delta_l)$ of these pairing channels, which are sketched in
Figure ~\ref{pairingstructure}, can be defined as follows,
\begin{eqnarray}
f_{{\rm NN},s} ({{\delta}_l})=1,
\nonumber \\
f_{{\rm NN},d+id} ({{\delta}_l})=e^{i(l-1)\frac {2\pi}{3}},
\nonumber \\
f_{{\rm NN},p+ip} ({{\delta}_l})=e^{i(l-1)\frac {2\pi}{3}},
\nonumber \\
f_{{\rm NNN}, p+ip} ({{\delta}_{l^{'}}})=e^{i(l^{'}-1)\frac {\pi}{3}},
\nonumber \\
f_{{\rm NNN}, f} ({{\delta}_{l^{'}}})=e^{i\frac{1+(-1)^{l^{'}}}{2}\pi},
\label{formfactor}
\end{eqnarray}
where the vectors ${\delta}_l$ ($l=1,2,3$) denote the NN inter-sublattice lattice
directions while $\delta_{l^{'}} $($l^{'}=1,2,3,4,5,6$) denote NNN intra-sublattice lattice directions.

For comparison purposes, we calculate the long-range averaged pairing correlation as
\begin{equation}
\overline{C_{\alpha}}(R>3) = \frac{1}{N_{1}}\sum_{R>3}C_{\alpha}(R),
\end{equation}
where $R$ is the distance between the electron pairs and $N_{1}$ is the number of $C_{\alpha}(R)$ with $R>3$. Here we ignore the contributions of short-distance pairing correlations ($R<3$) that mainly come from local spin and/or charge components.

Considering that the pairing correlation function may be affected by the non-interaction part\cite{Ying2018,White1989}. we also calculate the effective pairing correlation function, which can be defined as
\begin{equation}
V_{\alpha}(R)=C_{\alpha}(R)-\tilde{C_{\alpha}}(R),
\end{equation}
where $\tilde{C_{\alpha}}(R)$ is an uncorelated single-particle contribution, which can be obtained by the direct replace $\langle c^{\dagger}_{i\downarrow}c_{j\downarrow}c^{\dagger}_{k\uparrow}c_{l\uparrow}\rangle$ operator with  decoupled form $\langle c^{\dagger}_{i\downarrow}c_{j\downarrow}\rangle \langle c^{\dagger}_{k\uparrow}c_{l\uparrow} \rangle$ operator. We can determine the pairing channel $\alpha$ by the enhanced (suppressed) tendency of effective pairing correlation function.

The long-range effective pairing correlation function is also expressed as
\begin{equation}
\overline{V_{\alpha}}(R>3) = \frac{1}{N_{1}}\sum_{R>3}V_{\alpha}(R).
\end{equation}

We study the pairing correlation function by using CPQMC method, which is a sign-problem-free
auxiliary-field quantum Monte Carlo method and projects out the ground state
from a trial state by branching random walkers in the Slater determinant space.
 A constrained-path approximation is adoped to prevent the sign problem \cite{Zhang1997a,Zhang1997b,Huang2001a,Huang2001b,Liu2014,Liu2016}.
In our simulations,  we set the average number of random walkers to be
6000 and the time step $\Delta\tau=0.04$. 2000 Monte Carlo
steps were sampled before measurements, and 10 blocks of 300 Monte Carlo steps
were used to ensure statistical independence during the measurements. Closed-shell fillings were chosen in the simulations.

\section{Results}
\begin{figure}
\centering
\includegraphics[scale=0.2]{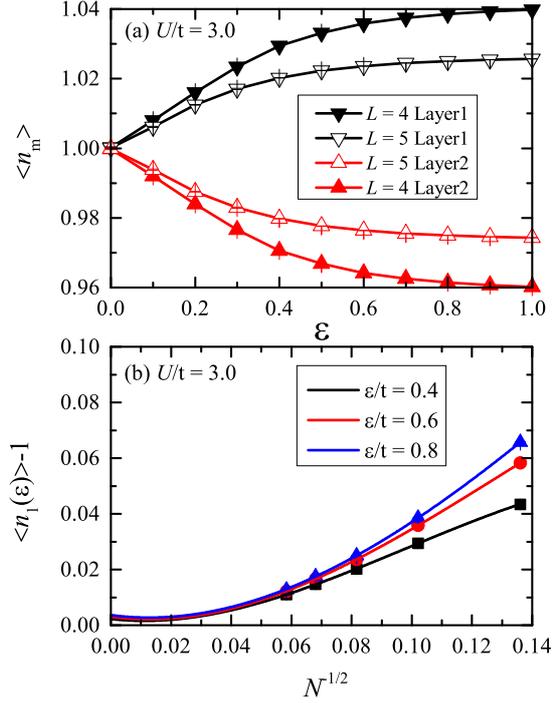}
\caption{(color online)
(a) Averaged electron density of each layer $\langle n_{m}\rangle$
versus potential difference $\varepsilon$ for
different lattice sizes. The filled symbols  and open symbols correspond to the lattice $L=4$ and $L=5$, respectively. (b) Effective electron doping density $\langle n_{1}(\varepsilon)\rangle-1$ of layer $1$ for various potential difference $\varepsilon$ using $3$rd order polynomials in $1/\sqrt{N}$, where $N$ is the each layer atom number of the system and $N = 6L^{2}$ $(L=3,4,5,6,7)$.
}\label{density}
\end{figure}

\subsection{Impact of electric field on electron density}
Since there exists a potential difference between the two layers when we turn on the external perpendicular electric field, it is natural that the distribution of electron density on each layer will change. As shown in Figure ~\ref{density}(a), the averaged electron density on each layer $\langle n_{m}\rangle$ is calculated on various $\varepsilon$. It is clear that $\varepsilon$ breaks the layer symmetry and more electrons transfer to the layer $1$ with perpendicular electric filed. Comparing the two lattices conditions of $L = 4$ and $L = 5$, we can obtain two important conclusions. Firstly, the electron density tends to increase gradually when $\varepsilon < 0.8t$ and approaches saturates when $\varepsilon > 0.8t$. The changing behavior of electron density can be explained as follows. Both on-site Hubbard repulsive interaction $U$ and potential difference $\varepsilon$ affect the transfer of interlayer charge. The Hubbard $U$ reflects the electron localization, which shows the effect of preventing the transfer of charge. However, potential difference $\varepsilon$ is beneficial to charge transfer between layers. The charge concentration saturates is the result of the competition between Hubbard $U$ and bias $\varepsilon$. Secondly, the effect of the electric field on the charge density is suppressed as the lattice size increases. To considering the effect of lattice size on charge density, we have shown the variation of the effective charge doping concentration $\langle n_{1}(\varepsilon)\rangle-1$ for different lattice sizes in the Figure ~\ref{density}(b). When $\varepsilon=0.8$, the fitted curves intercept with the vertical axis at small positive values $\sim 0.004$, indicating that charge density can be regulated by electric filed in the thermodynamic limit $(1/\sqrt{N}\to 0)$. These results demonstrate that the electric filed is an effective way to adjust the charge density of the AA-stacked bilayer graphene system.
In addition, the charge density in the experiment is usually dependent on the electric field. However, the explicit charge doping concentration is also related to the material and design of substrate and electrodes.
\begin{figure}
\centering
\includegraphics[scale=0.2]{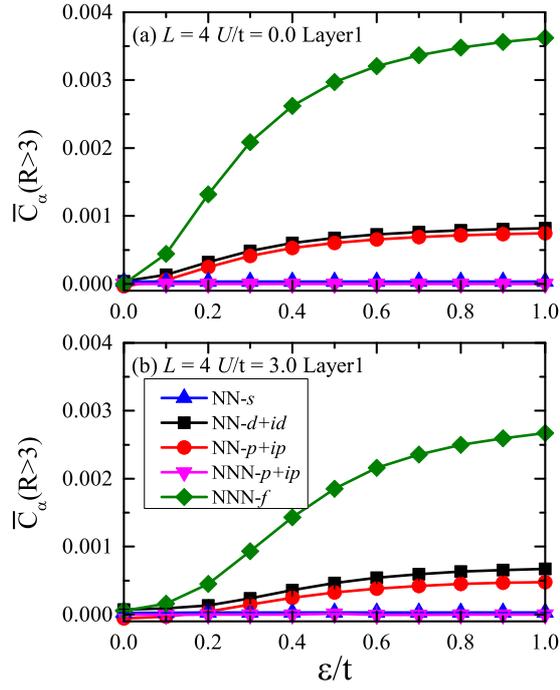}
\caption{(color online)
Long-range averaged pairing correlations of different pairing channels versus potential difference $\varepsilon$ on different Hubbard $U$ for the lattice size $L = 4$ (a) $U/t = 0.0$ and (b) $U/t = 3.0$.
}\label{pairingcorr}
\end{figure}

\begin{figure}
\centering
\includegraphics[scale=0.2]{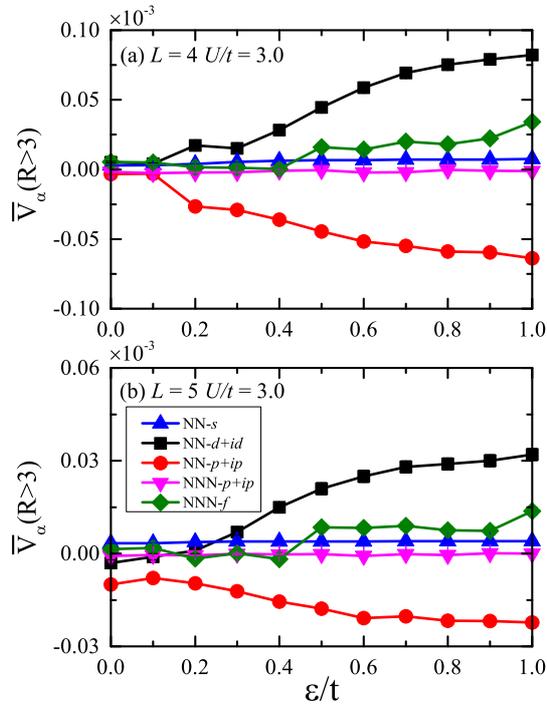}
\caption{(color online)
Long-distance averaged effective pairing correlations of different pairing channels versus potential difference $\varepsilon$ at $U/t = 3.0$ on different lattice sizes (a) $L = 4$ and (b) $L = 5$.
}\label{vertex}
\end{figure}

\subsection{Impact of electric field on electron correlations}
\begin{figure}
\centering
\includegraphics[scale=0.2]{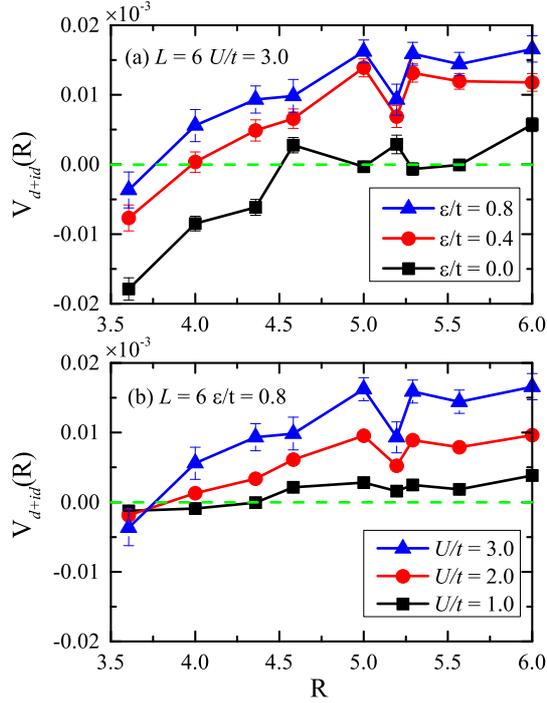}
\caption{(color online)
(a) Long-distance effective pairing correlation of $d+id$ pairing channels versus  distance $R$ between pairs on the lattice size $L = 6$ for different potential difference $\varepsilon$ at $U/t = 3.0$ and (b) different on-site interaction $U$ at $\varepsilon/t=0.8$. The green dash lines represent the position of 0.0.
}\label{vertexu}
\end{figure}
 Firstly, the impact of potential difference $\varepsilon$ on various pairing channels is discussed at half filling and lattice size $L = 4$ for different on-site Hubbard repulsive interaction $U/t = 0.0$ and $U/t = 3.0$. As shown in Figure ~\ref{pairingcorr}. It can be clearly seen that as we gradually increase the $\varepsilon$, the pairing correlation with triplet NNN-$f$ pairing channel responses sharply, which is much larger than other ones. Therefore, our simulation results seem to support that the system is triplet NNN-$f$ under the control of perpendicular electric field. However, considering that the superconducting pairing correlation is induced by the Hubbard interaction, the contribution of noninteracting part ($U/t = 0.0$) to the pairing correlation is meaningless and it might result in the misleading for pairing symmetry \cite{Ying2018,White1989}. The long-range averaged pairing correlations, as shown in Figure ~\ref{pairingcorr}, reflect that the NNN-$f$ pairing channel exceeds the NN-$d+id$ pairing channel, either interaction $U/t = 0.0$ or interaction $U/t = 3.0$. So, it can not be ruled out whether the dominance of NNN-$f$ pairing channels is attributed to the contribution of noninteracting. Furthermore, by comparing Figure ~\ref{pairingcorr}(a) and Figure ~\ref{pairingcorr}(b), It is evident that the shapes of the various pairing channels are very similar. This comparison indicates that the amplitude of pairing correlation of NNN-$f$ pairing channels may be much larger than other candidate channels, which is caused only by the electronic structure of $U/t = 0.0$, not by the effective attraction between electrons. Therefore, it is not possible to accurately determine the pairing form in the system from the point of view of the traditional pairing correlation function. The correlation of the interacting part of the pairing correlation function is the key to determine the form of electron pairing.
\begin{figure}
\centering\label{bandstructure}
\includegraphics[scale=0.3]{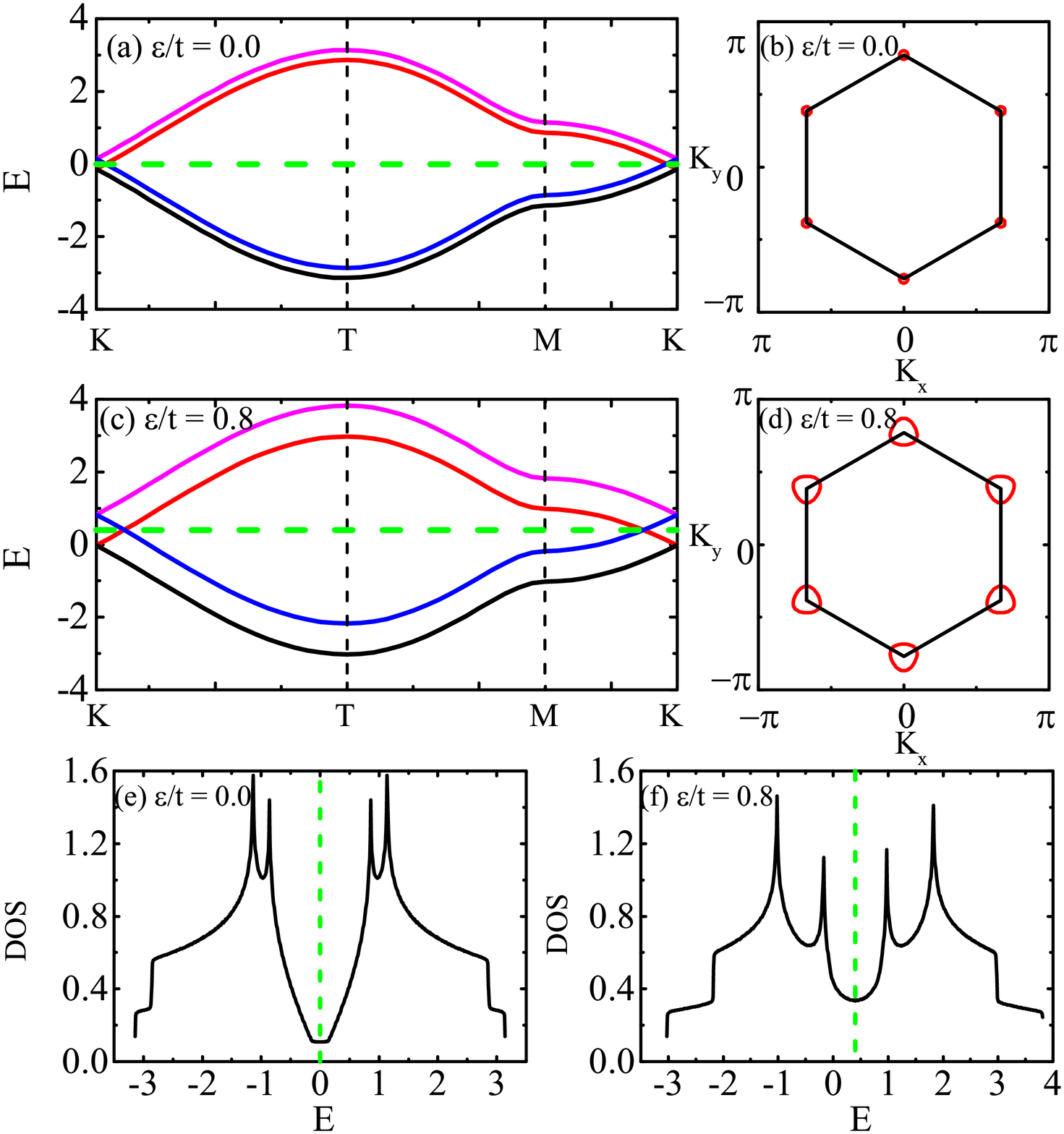}
\caption{(color online)
Band structures, Fermi surfaces and density of states of the non-interacting
Hubbard model on the AA-stacked bilayer honeycomb lattice on various potential difference $\varepsilon$. The green dash lines represent the position of Fermi levels.}\label{band}
\end{figure}

 Based on above discussion, we study the electron pairing by the effective pairing correlation functions versus various potential difference $\varepsilon$ in Figure ~\ref{vertex}. For NN-$s$ and NNN-$p+ip$ symmetries, the influence of the perpendicular electric field on their corresponding effective pairing correlation functions is little and the value is still close to $0.0$. For NN-$p+ip$ symmetry, the effective pairing correlation function gradually decreases with the change of the electric field potential and the value is negative. This shows that it is difficult to form an effective pairing attraction for the above three pairing channels. However, the values of the effective pairing correlation functions of NN-$d+id$ and NNN-$f$ are positive values and the correlation strength is gradually increased under the influence of the perpendicular electric field, which indicates that turning on $U$ and electric field produces effective pairing attraction in the NN-$d+id$ and NNN-$f$ pairing channels. Moreover, The effective pairing correlation function of NN-$d+id$ symmetry is much larger than the amplitude of NNN-$f$ symmetry, indicating that NN-$d+id$ is the dominant pairing channel in the AA-stacked bilayer graphene system under the perpendicular electric field.

 In order to clearly demonstrate the impact of electric field and on-site Hubbard repulsive interaction $U$ on the dominant $d+id$-wave, we also plotted the effective pairing correlation as a function of pairing range on the lattice size $L = 6$ in Figure ~\ref{vertexu}. The results of Figure ~\ref{vertexu}(a) demonstrate that the long range order of $d$-wave pairing symmetry increases when the electric field is turned on at all distances. Besides, our results show that the on-site coulomb interaction $U$ is beneficial to the pairing long range order, which can be easily obtained from Figure ~\ref{vertexu}(b).

\begin{figure}
\centering
\includegraphics[scale=0.2]{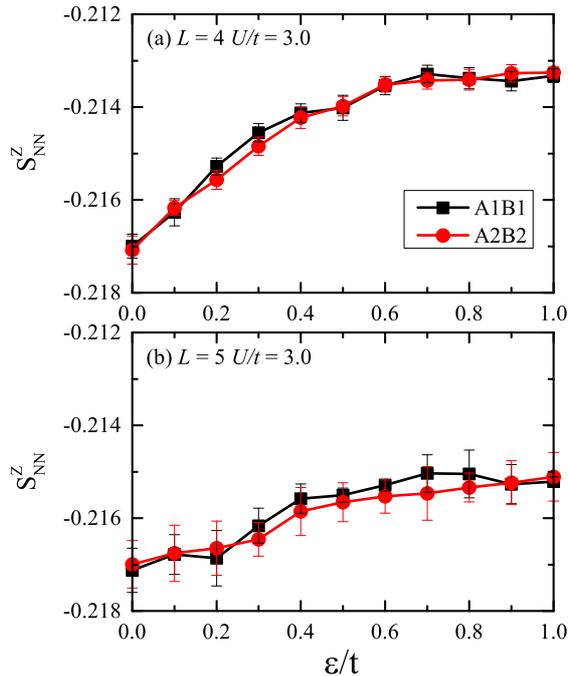}
\caption{(color online)
The intralayer NN spin correlation function versus potential difference $\varepsilon$
at $U/t = 3.0$ for different lattice sizes.(a) $L = 4$ and (b) $L = 5$.
}\label{spincorr}
\end{figure}

\subsection{ Pairing mechanism analysis}
Why $d+id$-wave is the dominant pairing channel after introducing the electron field?  In order to clarify this physical mechanism, we discussed the changes of band structure and spin correlations upon electric field. Firstly, we investigate the impact of electric field on the band structures, Fermi surface and density of states.
When the potential difference $\varepsilon/t = 0.0$, as shown in the top panels of Figure ~\ref{band}, the bands of each layer are almost degenerate and
the areas of Fermi surfaces are very limited. As potential difference $\varepsilon/t = 0.8$, the bands are further shifted and it makes the flat band around between M and K closer to the Fermi surfaces and enhances the density of states
near the Fermi level. Naturally, we speculate that an increase in the density of states is a key factor in favor of electron pairing.

Next we discuss the impact of electric field on magnetic properties of the
system. The intralayer NN spin correlation function
$S^{z}_{\rm NN}=\sum_{i,j=i+\delta_{l}}\langle (n_{i,m,\uparrow}-n_{i,m,\downarrow})(n_{j,m,\uparrow}-n_{j,m,\downarrow}) \rangle$ is calculated as a function of the potential difference $\varepsilon$ at half filling
and $U/t = 3.0$. From Figure ~\ref{spincorr}, the negative NN spin correlation function indicates that the system is an antiferromagnetic spin correlation at $\varepsilon/t = 0.0$. As we increase potential difference $\varepsilon$, one can see that the amplitude of spin correlations are suppressed. More specifically, comparing the NN antiferromagnetic spin correlation function at $\varepsilon/t = 0.0$ and $\varepsilon/t = 1.0$, the relative reduction of spin correlation function is very small for the lattice sizes $L = 4$ and $L = 5$, respectively.  Therefore, we can infer that AA stacked bilayer graphene can exhibit superconductivity induced by the antiferromagnetic spin fluctuations under the influence of electric field.

\section{Conclusions}
Understanding the pairing mechanism of unconventional superconductors has always been an important research topic. In this paper, we studied the impact of electric field on pairing properties and magnetic of the Hubbard model on the AA-stacked bilayer graphene. As electric field is turned on, a dominant chiral $d+id$-wave pairing is significantly enhanced. We also studied the band structures and spin correlations versus electric field. An increased density of states near Fermi surfaces and suppressed of antiferromagnetic spin correlations were discovered upon the electric field, which are the main trigger for the dominance of chiral $d+id$-wave pairing. Our simulations suggest the AA-stacked bilayer graphene is a candidate for chiral $d+id$ superconducting state under the control of electric field and our research provides an inportrant theoretical ideas for controlling the superconducting state on the bilayer honeycomb lattice.

\section*{Acknowledgment}
The authors thank Yongzheng Wu for insightful discussions. This work was supported by the National Natural Science Foundation of China (Grant No.~11674087).

\section*{References}


\end{document}